\documentstyle[multicol,aps,psfig]{revtex}
\renewcommand{\narrowtext}{\begin{multicols}{2}
\global\columnwidth20.5pc\noindent}
\renewcommand{\widetext}{\end{multicols}
\global\columnwidth42.5pc}
\def\boldc{\mbox{\boldmath$c$}}
\def\boldQ{\mbox{\boldmath$Q$}}
\def\boldq{\mbox{\boldmath$q$}}
\def\btau{\mbox{\boldmath$\tau$}}
\begin{document}
\draft
\preprint{September 2001}

\title{Continuum spin excitations in S=1 one-dimensional
antiferromagnet.
}
\author{Igor A. Zaliznyak}
\address
{Department of Physics, Brookhaven National Laboratory, Upton,
 New York 11973-5000}
\date{\today}

\maketitle

\begin{abstract}
We present detailed measurement of the non-hydrodynamic part of the spin
excitation spectrum in the model quasi-1D S=1 antiferromagnet CsNiCl$_3$
by inelastic magnetic neutron scattering. In the better part of the
Brillouin zone the effect of the inter-chain coupling is negligible, and
spin dynamics reflects that of a single Haldane chain. We find that at
$q\lesssim 0.6\pi$ this quantum spin system ceases to support coherent
propagating excitation, which gradually turns into a continuum band of
states, whose width grows with decreasing $q$. This finding is
consistently verified under different resolution conditions, obtained
with two different high-luminosity experimental setups.

\end{abstract}
\pacs{PACS numbers:
       75.10.Jm,  
       75.40.Gb,  
       75.50.Ee}  
\narrowtext

\section{Introduction}

In striking neglect of the na\"{\i}ve ``common wisdom'', Haldane ground
state (GS) of the one-dimensional (1D) S=1 Heisenberg antiferromagnet
(HAFM)
\cite{Haldane,Takahashi,MeshkovDeisz,White,Sorensen,Regnault,Broholm,Buyers,Steiner,Zaliznyak1994}
does not connect the Neel-ordered ``spin solid'' GS of the semiclassical
S$\gg 1$ HAFM with the almost ordered ``marginal liquid'' state of the
S=1/2 chain \cite{Bethe}. Instead, it is a ``quantum liquid'' with
finite correlation length and a gap in the spin excitation spectrum.
Spectral weight of the spin fluctuations is concentrated in a long-lived
massive triplet mode in the neighborhood of the Brillouin zone (BZ)
boundary $q=\pi$. Any remainder of the spectacular continuum observed at
$q\approx\pi$ in the S=1/2 1D HAFM \cite{Dender} is predicted to be
extremely faint \cite{Takahashi,MeshkovDeisz,Horton}. On the other hand,
both nonlinear $\sigma$-model, which is believed to be a valid
description of the S=1 HAFM chain in the long-wavelength limit, and the
variational treatment based on the Jordan-Wigner fermionization, predict
that two-magnon continuum states are the lowest-energy excitations at
$q\approx 0$ and dominate the spectrum in the vicinity of the Brilloun
zone center \cite{AffleckWeston,GomezSantos}. In the absence of magnon
interaction the two-magnon continuum starts above a threshold energy
$\varepsilon_{2m}(q) = min\{2\cdot \varepsilon(\pi+q/2),\; \varepsilon
(\pi+q) +\Delta_H\}$, smallest at $q=0$, where it is $2\Delta_H$, twice
the Haldane gap.

Experimental observation of the continuum part of spin excitation
spectrum in Haldane chain is a very challenging problem. Primarily this
is due to the rapid decrease of the static structure factor $S(q)$,
which gives the energy-integrated intensity of the scattering
cross-section by spin fluctuations, at small $q$
\cite{Takahashi,MeshkovDeisz,White,Broholm}. In single-mode
approximation (SMA) $S(q)\sim (1-\cos{\rm q})/E(q)$ vanishes $\sim q^2$
as $q\rightarrow 0$. In fact, in the state of the art neutron scattering
experiment \cite{Broholm} authors found no sign of a continuum in the
model Haldane chain compound NENP down to $q=0.3\pi$. This result has
lead to a widespread belief that there is indeed no continuum at $q >
0.3\pi$, despite indications to the contrary by MC studies
\cite{MeshkovDeisz,White,Sorensen}. However, upon careful examination
the findings of the Ref. \onlinecite{Broholm} are not at all that
prohibitive. To merely observe magnetic scattering at $q=0.3\pi$ authors
employed very coarse resolution, with full width at half maximum (FWHM)
of the instrument wavevector acceptance covering about a quarter of the
1D BZ. Their statement that measured magnetic intensity is consistent
with the SMA cross-section, treated within the context of this broad
resolution, only imposes an upper limit, and not a very stringent one,
on the width of the continuum. Another obstacle to observing the
continuum magnetic scattering is a single-ion anisotropy in the spin
Hamiltonian. As in the majority of Ni-organic chain compounds, it is
quite large in NENP, and results in a significant splitting of spin
fluctuations with different polarizations. Smeared by the resolution
such splitting is hard to distinguish from continuum.

Quasi-1D antiferromagnet CsNiCl$_3$ is one of the most isotropic
and best studied Haldane model compounds. It provided some of the
earliest, although initially controversial, experimental evidence
in favor of the Haldane conjecture \cite{Buyers}. In CsNiCl$_3$
super-critical inter-chain exchange coupling $J'\approx 0.03J$
results in a long-range order below $T_N\approx 4.8$~K, but as
temperature rises above $T_N$, a gap opens in the spin excitation
spectrum, and it quickly recovers properties of an isolated S=1
HAFM chain \cite{Steiner,Zaliznyak1994}. Unfortunately,
characterization of  $T=0$ continuum is impossible at such
elevated temperatures because of the substantial magnon thermal
broadening.
However, it was recently pointed out on the basis of the MF-RPA
(mean field random phase approximation) analysis that inter-chain
coupling modifies only the low-energy part of the excitation
spectrum, and therefore even at $T<T_N$ dynamic spin response of
CsNiCl$_3$ in the better part of the BZ around the top of the 1D
dispersion is identical to that of an individual chain
\cite{Zaliznyak2001}. Here we present neutron scattering
measurements which illustrate this point and reveal the existence
and the extent of the excitation continuum.

\section{Experiment}

We studied large $ m\approx 6.4$~g sample of CsNiCl$_3$, composed
of two single crystals, co-aligned to yield effective mosaic
$\lesssim 1^\circ$. Sample was mounted on an Al plate in the
standard ``ILL orange'' 70 mm cryostat with $(h,h,l)$ zone in the
scattering plane. CsNiCl$_3$ has hexagonal structure $P63/mmc$
with two equivalent ions per $c$ spacing, so that $\boldQ=(h,k,l)$
in reciprocal lattice units (rlu) corresponds to $q_\parallel =
\pi l $ in the 1D BZ of a chain. Sample had longer dimension
parallel to the hexagonal \boldc -axis, and was in the
transmission ``Laue'' geometry shown in Fig. 1. Magnetic
scattering was measured at the ``base'' temperature $T=1.5(2)$~K,
the non-magnetic background (BG) was collected in identical scans
at $T=150$~K.

Experiments were performed on SPINS 3-axis cold neutron spectrometer at
NIST Center for Neutron Research, using two complementary setups
illustrated in Figure 1 (a), (b).
Monochromatic incident neutron beam was obtained by (002) reflection
from vertically focused pyrolytic graphite (PG) monochromator, viewing
the $^{58}$Ni neutron guide. While in setup (a) the ``natural'' beam
collimation around the sample was $\approx 100'-540'$, it was restricted
to $80'-80'$ by Soller collimators in setup (b). Scattered neutron
wavevector was analyzed by (002) reflection from an array of 11 (a) or 9
(b) independently rotating $\approx 2$ cm wide PG crystals through a
$80'$ radial collimator (RC) onto $\approx 24.3$ cm wide position
sensitive detector (PSD) with 256 pixels. Calibration of the scattered
neutron energy accepted by PSD pixel and its sensitivity was done by
measuring elastic incoherent scattering from standard sample at
different incident energies.

\begin{figure}[htb]
\label{fig1}
\begin{center}
\mbox{\psfig{figure=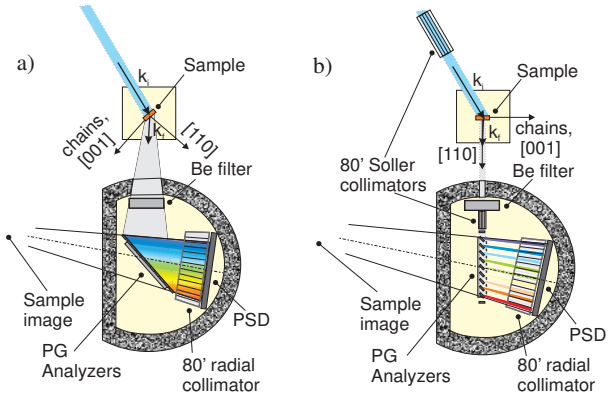,width=8cm,angle=0}}
\end{center}
\caption{Two high count rate setups with PSD used for measurement
of the 1D spin excitations in CsNiCl$_3$. (a) large flat PG
analyzer gives about 10-fold increase in acceptance to the
direction of the scattered neutrons. (b) 9 PG segments aligned to
create a polychromatic image of the sample at the RC center
similarly increase acceptance to the length of ${\bf k}_f$.}
\end{figure}

Use of the high count rate setups which employ SPINS large area
segmented PG analyzer and matching large PSD was key to the success of
our measurements. Essential element in such setup is a radial collimator
in front of the PSD. It dramatically restricts acceptance of the PSD
pixels to incoherent scattering from the analyzer and its support
structures, which otherwise render phohibitively large background. RC
requires that reflection from the analyzer produce a geometrical image
of the sample at its curvature center. For each incident wavevector
${\bf k}_i$ the direction and/or length of the scattered wavevector
${\bf k}_f$ are functions of the pixel position across PSD, as defined
by geometry of the Bragg reflection at the analyzer, Fig. 1. This
results in a coupled scan with both energy transfer $E$ and wavevector
\boldq\ in the sample reciprocal space varying across the PSD. However,
\boldq\ component along dispersive direction of interest (chain in our
case) can be kept $q_\parallel \approx const$ in such scan, if this
direction is aligned perpendicular to the direction of the scan by
appropriately choosing $q_\perp$, as shown in Fig. 1.

{\it Setup (a)}. PSD central energy was fixed at
$E_f^{(0)}=4.2$~meV, and range $E_f \in [3.6, 4.8]$~meV was
covered in this measurement. Evidently, large flat analyzer simply
creates a (polychromatic) mirror image of the sample. Radius of
the SPINS RC is equal to the sample-detector flight path,
optimizing this setup.  For all PSD pixels (all sample scattering
angles $2\theta_s$) ${\bf k}_f$ component along the analyzer Bragg
wavevector \btau$_A$ is constant, $=\tau_A$/2. Therefore, in the
paraxial approximation constant is $q_\parallel\approx \frac{({\bf
k}_i \btau_A)}{\tau_A} - \frac{\tau_A}{2}$, while $q_\perp$ varies
across PSD as a function of ${\bf k}_f$.

{\it Setup (b)}. PSD central energy was fixed at $E_f^{(0)}=4.57$~meV.
Angle between the consecutive PG segments $\delta(\theta_A) \approx
35'\div 40'$, and the corresponding difference in the reflected neutron
energy, were chosen to create a polychromatic image of the sample at the
RC center. 9 analyzer segments provided full PSD coverage, reflecting
energies in the range $E_f \in [4.03, 5.13]$~meV. Chain direction was
aligned perpendicular to ${\bf k}_f$, and therefore, for given ${\bf
k}_i$ and $2\theta_s$, only $q_\perp$ varied across PSD, while
$q_\parallel = k_i\sin 2\theta_s = const$.
Important feature of this setup is a possibility to restrict analyzer
angular acceptance by inserting a collimator after the sample. This
allows measurement at smaller scattering angles, and also
shapes/tightens the instrument resolution function, as shown by the
resolution FWHM ellipses in Fig. 2.

\section{Results and discussion}

In Figure 2 we show contour plots of the measured spectral density
of the magnetic scattering intensity, $I(q,E)/\int I(q,E)dE$, with
the linear $(q,E)$-dependent background subtracted. It is evident
in both panels that spectrum acquires finite width in energy at $l
\lesssim 0.5$.

Fig 2 is constructed from the raw data (integration is done via point by
point summation), and is slightly distorted by the instrument
resolution. In the single-mode part of the spectrum, at $l \gtrsim 0.6$,
measured line-shape is completely defined by the interplay of dispersion
and resolution. Although resolution volume is smaller in setup (b), the
``focusing'' effect (longer axis of the FWHM ellipse is parallel to the
dispersion in Fig. 2(a)) results in sharper peaks in setup (a). In
principle, an opposite ``de-focusing'' effect is of concern for $l
\lesssim 0.5$ measurements in this setup, as it would result in quite
significant broadening even of a single-mode spectrum. However, careful
accounting for the resolution shows that non-zero {\it intrinsic} width
at $l \lesssim 0.5$ accounts for $\gtrsim 2/3$ of the spectral width
measured in setup (a). ``De-focusing'' is absent in setup (b), where the
FWHM ellipse is approximately round, Fig 2(b). Spectacular agreement of
the $q_\parallel \lesssim 0.5\pi$ spectrum measured in two setups shows
explicitly, that observed cross-over from a single mode to continuum
with significant intrinsic width is not a resolution effect.

\begin{figure}[htb]
 \label{fig2}
\begin{center}
\mbox{\psfig{figure=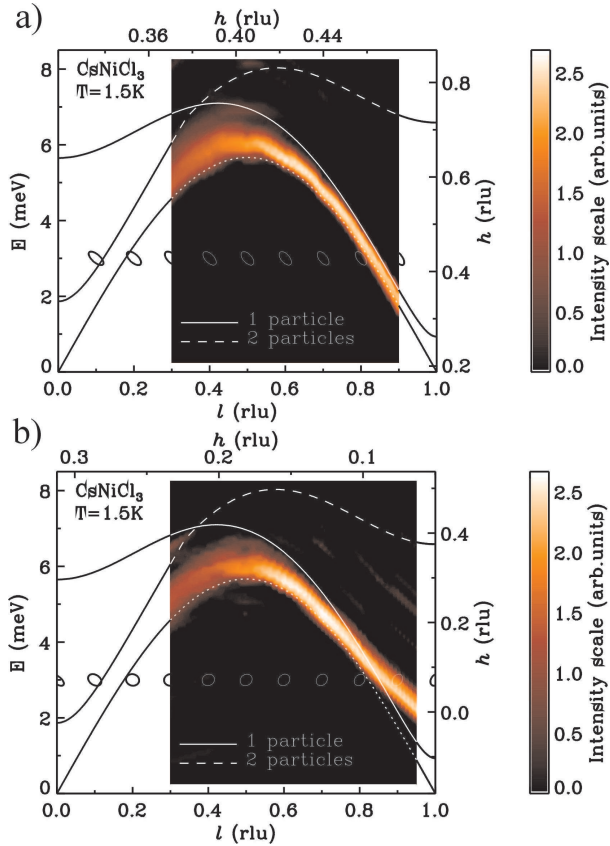,width=8cm,angle=0}}
\end{center}
\caption{Contour plot of the spectral density of magnetic scattering
reconstructed from the constant-$q_\parallel$ scans, measured in setups
(a) and (b) of Fig.~1 correspondingly, via linear interpolation. Scale
on the right shows variation with energy of the wavevector transfer
perpendicular to the chain at $l=0.5$, scale on the top -- its variation
with $l$ at $E=3$~meV. Ellipses are calculated half maximum contours of
the instrument resolution function at $E=3$~meV. Solid curve is the
single-magnon dispersion discussed in text, dashed line shows the lowest
energy of two non-interacting magnons with given total $q_\parallel=\pi
l$, dotted line is $\varepsilon(q_\parallel)= 2.49J \sin{q_\parallel}$.
}
\end{figure}

Another important distinction between the data in Fig 2(a) and (b), is
that $q_\parallel\approx const$  geometry imposes different conditions
on $q_\perp$ in setups (a) and (b) correspondingly. Therefore,
measurements are done along different trajectories in the sample 3D
reciprocal space, as illustrated by the right and top axes in Fig 2. The
effect of the inter-chain coupling is manifested at low energies by
difference in dispersion at $l \gtrsim 0.8$. While in (a) $h \sim 0.3$
is close to the 3D magnetic Bragg position at $l \approx 1$, for setup
(b) it is close to the top of inter-chain dispersion, $h\approx 0$. In
agreement with MF-RPA analysis, effect of the inter-chain coupling
becomes insignificant at $0.25 \lesssim l \lesssim 0.75$. For
comparison, curves in Fig 2 show magnon spectrum in a Haldane chain
obtained in \cite{GomezSantos}, for $J=2.275$ meV, $\Delta_H=0.41J \;, v
= 2.49 J$, $\alpha = v$ \cite{Zaliznyak2001}.


Our results show explicitly and unambiguously, that single-mode
excitation becomes unstable around the top of the dispersion band in a
Haldane chain. Instead, a continuum excitation spectrum, whose width
increases with decreasing $q$, is observed at $q \lesssim 0.6\pi$. In
fact, very similar behavior was found in another quantum liquid --
superfluid $^4$He, where the ``maxon'' excitation turns into a broad
continuum-like feature under pressure \cite{Passell}, which suppresses
the ``roton'' gap and drives the system towards crystallization, a
quantum phase transition.

It is my pleasure to thank C. Broholm, who introduced me to the
concept of PSD techniques and shared many invaluable ideas, and
S.-H. Lee, who has contributed greatly to the success of this
measurement. This work was carried out under Contract
DE-AC02-98CH10886, Division of Materials Sciences, US Department
of Energy. The work on SPINS was supported by NSF through
DMR-9986442.

\widetext

\begin{references}

\bibitem{Haldane}
 F.~D.~M.~Haldane, Phys. Lett. {\bf 93A}, 464 (1993);
 Phys. Rev. Lett. {\bf 50}, 1153 (1983).

\bibitem{Takahashi}
 M.~Takahashi, Phys. Rev. Lett. {\bf 62}, 2313 (1989);
 Phys. Rev. B {\bf 50}, 3045 (1994).

\bibitem{MeshkovDeisz}
 S.~V.~Meshkov, Phys. Rev. B {\bf 48}, 6167 (1993);
 J.~Deisz, M.~Jarrell, D.~L.~Cox, Phys. Rev. B {\bf 48}, 10227 (1993).

\bibitem{White}
 S.~R.~White, Phys. Rev. Lett. {\bf 69}, 2863 (1992);
 S.~R.~White and D.~A.~Huse, Phys. Rev. B {\bf 48}, 3844 (1993).

\bibitem{Sorensen}
 E.~S.~Sorensen, I.~Affleck, Phys. Rev. B {\bf 49}, 13235, (1994);
 Phys. Rev. B {\bf 49}, 15771 (1994).

\bibitem{Regnault}
 J.-P.~Renard {\it et al }, Europhys. Lett. {\bf 3}, 949 (1987);
 L.-P.~Regnault {\it et al }, Phys. Rev. B {\bf 50}, 9174 (1994).

\bibitem{Broholm}
 S.~Ma {\it et al }, Phys. Rev. Lett. {\bf 69}, 3571 (1992).

\bibitem{Buyers}
 W.~J.~L.~Buyers {\it et al }, Phys. Rev. Lett. 56, 371 (1986);
 R.~M.~Morra {\it et al }, Phys. Rev. B {\bf 38}, 543 (1988);
 Z.~Tun {\it et al }, Phys. Rev. B {\bf 42}, 4677 (1990).

\bibitem{Steiner}
 M.~Steiner {\it et al }, J. Appl. Phys. {\bf 61}, 3953 (1987);

\bibitem{Zaliznyak1994}
 I.~A.~Zaliznyak {\it et al }, Phys. Rev. B {\bf 50}, 15824 (1994).

\bibitem{Bethe}
 H.~Bethe, Z. Phys. {\bf 31}, 205 (1931).

\bibitem{Dender}
 D.~C.~Dender {\it et al }, Phys. Rev. B {\bf 53}, 2583 (1996).

\bibitem{Horton}
 P.~Horton and I.~Affleck, cond-mat/9907431.

\bibitem{AffleckWeston}
 I.~Affleck, R.~A.~Weston, Phys. Rev. B {\bf 45}, 4667 (1992).

\bibitem{GomezSantos}
 G.~G\'{o}mez-Santos, Phys. Rev. Lett. {\bf 63}, 790 (1989).

\bibitem{Zaliznyak2001}
 I.~A.~Zaliznyak {\it et al },
 Phys. Rev. Lett. {\bf 87}, 017202 (2001).

\bibitem{Passell}
E.~H.~Graf {\it et al },
Phys. Rev. A {\bf 10}, 1748 (1974).


\end{references}
\end{document}